\documentclass[conference,a4paper]{APSIPA2026}
\usepackage{amsmath}
\usepackage{graphicx}
\usepackage{multirow}
\usepackage{threeparttable}
\usepackage[backend=biber,style=ieee,]{biblatex}
\usepackage{geometry}
\usepackage{fancyhdr}

\fancypagestyle{firststyle}{
  \fancyhf{}
}

\begin{document}

\title{
EEG-Based Characterization of Samatha and Vipassana Meditation States 
}




\author{
\authorblockN{
M. A. B. C. A. Bandaranayake$^*$, K. P. U. Chandrathilake$^*$,  J. L. S. Jayasekara$^*$, S. T. Piyasena$^*$,
\\ A. T. L. K. Samrasinghe and
Wageesha N. Manamperi
}

\authorblockA{
Department of Electronic and Telecommunication Engineering, University of Moratuwa, Sri Lanka \\
E-mail: \textit{\{chamidianuththara, uththara56, lashinisharikaj, sajanit95\}@gmail.com, kithsiris@uom.lk,wageesham@uom.lk}}

}

\maketitle
\let\thefootnote\relax\footnotetext{$^*$These authors contributed equally to this work.}
\thispagestyle{firststyle}
\pagestyle{empty}

\begin{abstract}
Meditation has been associated with a range of cognitive and physiological benefits. However, the underlying neural mechanisms of different meditation practices are not yet fully understood. This study investigates whether electroencephalogram (EEG) signals can be used to characterize and distinguish two commonly practiced meditation techniques, namely Samatha and Vipassana, which involve distinct cognitive processes, with Samatha emphasizing sustained concentration and Vipassana emphasizing mindful observation and insight. By extracting features such as band power, coherence and wavelet entropy from EEG signals recorded during pre-meditation resting, Samatha, and Vipassana states, we provide an assessment of how spectral power, signal complexity, and functional connectivity differentiate these practices. 
Preliminary results from $12$ experienced meditators indicate condition-dependent EEG differences, with the most pronounced effects observed in the delta frequency band.
These results demonstrate the utility of EEG-based metrics for the objective characterization of meditation and advance our comprehension of the neural processes underlying different contemplative practices.
\end{abstract}


\vspace{-0.2cm}
\section{Introduction}
\label{sec:2}

The increasing popularity of meditation has led to the rapid integration of preliminary neuroscientific findings into consumer-based technologies, such as Electroencephalography (EEG) assisted neurofeedback devices for meditation \cite{brandmeyer2013meditation}. Proponents argue that reproducible EEG markers can be linked to specific contemplative practices, potentially aiding practitioners in reaching desired mental states. However, this commercial expansion is concerning, as neurofeedback can result in adverse effects when applied improperly without a comprehensive understanding of the underlying neural mechanisms \cite{hammond2008first}. In this paper, we address this issue by investigating two primary meditation practices, namely Samatha and Vipassana, based on their EEG signatures and providing a comprehensive analysis.

Previous studies have investigated a variety of meditation traditions, including mindfulness, Zen, transcendental meditation, yoga-based meditation, and Vipassana practices \cite{atad2025meditation}. Rooted in Buddhist tradition and widely practiced across Asia, Samatha and Vipassana represent two fundamental meditative approaches \cite{ray2004presence}. Samatha is a concentration-based practice focused on achieving mental calmness and sustained attention, while Vipassana is an insight-based practice centered on cultivating awareness through the direct observation of phenomena. Although both aim to enhance well-being, their distinct cognitive demands suggest divergent neural correlates.

Most studies utilized EEG to examine meditation-related changes in frequency band power, or entropy \cite{atad2025meditation,young2021contrasting,cahn2010occipital,vivot2020meditation}. More recently, efforts have shifted toward utilizing magnetoencephalography (MEG) \cite{pascarella2025meditation} and functional magnetic resonance imaging (fMRI) \cite{young2018impact} to map the cortical regions, connectivity patterns, and functional networks associated with meditative states. Despite the advancements provided by these modalities, EEG remains a vital tool for studying meditation due to its high temporal resolution and portability, facilitating the development of accessible, real-time assessment frameworks.
Nevertheless, comparative EEG research on these practices remains limited, particularly among experienced meditators, and is hindered by persistent methodological limitations, leaving the precise neurophysiological signatures of different meditative practices poorly understood \cite{atad2025meditation}.


In this study, we address this gap by characterizing Samatha and Vipassana meditation states through EEG signal analysis.
The main contributions of this work are threefold: (1) a comparative EEG analysis of Samatha, Vipassana, and resting states in experienced meditators; (2) a unified feature analysis using band power, coherence, and wavelet entropy; and (3) identification of frequency-specific and region-specific EEG features that may help distinguish these meditation states, with delta band effects emerging as the most prominent in this dataset.
Experimental results demonstrate the neurophysiological patterns associated with pre-meditation resting, Samatha, and Vipassana states across major EEG bands: delta, alpha, and gamma, which contribute to the development of an EEG-assisted neurofeedback device for assessing meditation states.

\vspace{-0.3cm}
\section{Problem Statement  }
\label{sec:2}

Meditation practices such as Samatha and Vipassana involve distinct cognitive processes. While Samatha meditation develops calmness and concentration through the suppression of the five hindrances, Vipassana meditation develops insight into the nature of ultimate realities, leading to an understanding of causality.
Owing to these differences in cognitive engagement, the two meditation practices are expected to exhibit distinct neural activity patterns.

Electroencephalography (EEG) provides a non-invasive means of measuring brain activity with high temporal resolution. 
Based on EEG signatures, this paper aims to investigate whether EEG signals can be used to objectively characterize Samatha and Vipassana meditation states, as elaborated in the next section.


\vspace{-0.2cm}
\section{Materials and methods }
\label{sec:3}
\vspace{-0.1cm}


In this section, we first describe the EEG data acquisition procedure, followed by the preprocessing steps applied to the recorded signals. The EEG signals are then decomposed into subbands using wavelet decomposition techniques \cite{percival2000wavelet}, and finally, EEG features are extracted.

\vspace{-0.2cm}
\subsection{EEG Acquisition and Participants }\label{sec:3.1}
\vspace{-0.1cm}

\subsubsection{Participants}
A total of $N=12$ meditators ($F=3$, $M=9$) participated in the study ($M= 53.42, SD=10.88,$ age range: $35- 75$
years). The participants had been practicing meditation for a considerable period
($M=15.41, SD=14.45,$ range: $2-35$ years), and reported meditating for at least one hour per day. Participants were recruited from the local meditation community through word-of-mouth referrals and telephone invitations.
Given the modest sample size and the broad range of participant age and meditation experience, the present study should be regarded as an exploratory analysis. 
The study received institutional ethics approval, and written informed consent was obtained from all participants.


\subsubsection{Recording Conditions}

EEG data were recorded using a 32-channel g.GAMMAcap (g.tec Medical Engineering GmbH) with electrode positions arranged according to the International 10–20 system with sampling rate of $256$ Hz. For regional analysis, the electrodes were grouped into frontal (Fp1, Fp2, AF3, AF4, F7, F3, Fz, F4, and F8), central (FC5, FC1, FC2, FC6, C3, Cz, and C4), parietal (CP5, CP1, CP2, CP6, P7, P3, Pz, P4, and P8), occipital (PO7, PO3, PO4, PO8, and Oz), and temporal (T7 and T8) regions. Impedance levels were kept below $50~$k$\Omega$.

\subsubsection{Procedure}

The participants were instructed to sit comfortably in a quiet environment and performed a brief calibration procedure by blinking a few times and keeping their eyes closed for several seconds to verify proper EEG signal acquisition. Each participant completed three experimental conditions: i) a pre-meditation resting state, ii) Samatha, and iii) Vipassana. Prior to each session, a short 2-minute stabilization period was provided during which participants were instructed to blink three times and then relax with eyes closed to ensure signal stability.

The experiment began with a pre-meditation resting condition, where participants remained in a relaxed, eyes-closed state without engaging in any meditative practice for a duration of $5$ minutes. Following this, participants were informed of the transition to the meditation state using a bell sound. Each participant then completed two meditation sessions: Samatha and Vipassana meditation, with a short break between sessions to reduce carryover effects. Note that the participants were asked to minimize unnecessary movements and maintain a relaxed posture throughout the recording period to reduce motion-related artifacts.

In this study, loving-kindness (\textit{metta}) meditation was selected as the representative Samatha practice. During meditation, practitioners first establish mindfulness (\textit{sathi}) toward themselves and all living beings and progressively cultivate feelings of loving-kindness. This practice leads progressively to an access concentration state characterized by reduced influence of mental hindrances and increased mental stability. With continued practice, participants were expected to reach the first concentration (\textit{dhyana}) state, where mental factors such as applied thought (\textit{vitakka}), sustained investigation (\textit{vicāra}), joy (\textit{pīti}), happiness (\textit{sukha}), and one-pointedness (\textit{ekaggatā}) are present. In deeper practice, participants may reach the second concentration state, where applied and sustained thought diminish, and the experience is dominated by joy, happiness, and one-pointedness of attention.

In the Vipassana practice, participants were instructed to maintain mindfulness (\textit{sathi}) on a selected object (e.g., body sensations or auditory perception) while observing its arising and passing away. The practice emphasized insight into impermanence (\textit{anicca}), suffering (\textit{dukkha}), and non-self (\textit{anattā}), along with related contemplative insights such as disenchantment (\textit{nibbidā}), dispassion (\textit{virāga}), cessation (\textit{nirodha}), and relinquishment (\textit{paṭinissagga}).

Each meditation session lasted approximately $20 - 30$ minutes. 
After each session, participants completed a post-experimental questionnaire to qualitatively validate their perceived meditation state and level of concentration.


\vspace{-0.2cm}
\subsection{EEG Preprocessing\label{sec:3.2}}
\vspace{-0.1cm}

The recorded EEG signals were preprocessed using the open-source EEGLAB toolbox \cite{delorme2004eeglab} to remove noise and artifacts and to obtain clean signals suitable for further analysis. The preprocessing pipeline consisted of several sequential stages, including filtering, channel rejection, interpolation, re-referencing, burst noise removal, and artifact removal.

\subsubsection{Filtering}
The recorded signals were filtered with a band
pass of $0.5 - 60$ Hz to remove slow baseline drifts and high-frequency noise components. The power-line interference of $50$ Hz was removed  using a notch filter.

\subsubsection{Channel Rejection, and Interpolation}

The noisy channels that contain more than $5\%–10\%$ corrupted data, were removed using an inbuilt EEGLAB automatic channel rejection function based on the
electrode joint probability kurtosis measure with a maximum Z-score threshold of $5$ \cite{delorme2004eeglab}. These excluded channels were subsequently reconstructed via spherical interpolation \cite{srinivasan1999methods} using the remaining clean channels.

\subsubsection{Average Referencing}

The filtered and interpolated EEG data were subsequently re-referenced to the global average of all electrodes. This average reference procedure assumes that net charge creation within the brain is zero, implying an absence of isolated monopolar sources and sinks.

\subsubsection{Burst Noise Removal}

To eliminate high-amplitude artifacts without compromising underlying physiological brain dynamics, burst noise was processed using the artifact subspace reconstruction (ASR) algorithm \cite{chang2018evaluation}. Data segments exhibiting a variance larger than a predefined threshold relative to the clean, calibrated data were identified and reconstructed. To ensure that genuine and useful signal information was not inadvertently lost during this preprocessing stage, a conservative and highly permissive variance threshold of 25 was applied. Consequently, the ASR function was strictly restricted to isolating and removing only extremely high-amplitude noise bursts.

\subsubsection{Artifact Removal}

In addition to channel-level filtering, a semi-automated analysis was performed using independent component analysis (ICA) with the aid of the multiple artifact removal algorithm (MARA) \cite{mannan2018identification}. MARA was used to estimate the probability of each component being an artifact based on spatial, spectral, and temporal features. Components associated with eye blinks, eye movements, muscle activity, and electrode noise were identified and removed following manual inspection of their scalp topographies, power spectra, and time-domain characteristics, together with the MARA output. This semi-automated approach ensured reliable artifact rejection while preserving neural components relevant to the subsequent frequency domain analysis, which we discuss next.

\vspace{-0.2cm}
\subsection{Band Decomposition
}\label{sec:3.3}
\vspace{-0.1cm}

The preprocessed EEG signals were decomposed into their constituent frequency bands using the maximal overlap discrete wavelet transform (MODWT) \cite{percival2000wavelet}. Compared to the conventional discrete wavelet transform (DWT), MODWT does not involve downsampling and provides improved time-frequency localization for non-stationary signals such as EEG. The 'Daubechies-8' (db8) wavelet was selected as the mother wavelet due to its suitability for EEG analysis. In this study, we focus on three frequency bands, delta ($0.5 – 4$ Hz), alpha ($8-12$ Hz), and gamma ($32 – 60$ Hz) bands.

\vspace{-0.2cm}
\subsection{Feature Computation
}\label{sec:3.4}
\vspace{-0.2cm}


\subsubsection{Band Power}
The power values of the recorded signals (in $\mu$V$^2$) were log-transformed and expressed dB units using $10 \times log_{10} (\mu$V$^2)$ formula to normalize the power value distributions prior to statistical analysis. Band power features were computed from the wavelet coefficients corresponding to the delta, theta, alpha, beta, and gamma frequency bands. The average power of the $k^{\mathrm{th}}$ frequency band was calculated as
\begin{equation}
P_k=\frac{1}{N}\sum_{j=1}^{N}|W_{k,j}|^2,
\end{equation}
where $W_{k,j}$ denotes the wavelet coefficient at the $k^{\mathrm{th}}$ decomposition level and $N$ is the number of samples. Band power features were extracted according to the regional, hemispheric, and frequency-band hierarchies.

\subsubsection{Coherence}
Coherence analysis was performed to quantify the functional connectivity between pairs of EEG electrodes located in different scalp regions. The magnitude-squared coherence between two EEG signals
recorded at electrodes $x$ and $y$, was given as
\begin{equation}
C_{xy}(f)=\frac{|P_{xy}(f)|^2}{P_{xx}(f)P_{yy}(f)},
\end{equation}
where $P_{xx}(f)$ and $P_{yy}(f)$ denote the power spectral densities of $x(t)$ and $y(t)$, respectively, and $P_{xy}(f)$ denotes their cross-power spectral density at frequency $f$. The coherence value ranges from $0$ to $1$, where larger values indicate stronger synchronization between the two EEG signals. Both intra-hemispheric and inter-hemispheric electrode pairs were considered in the analysis.


\subsubsection{Entropy}
Wavelet entropy was used to characterize the complexity and degree of order/disorder associated with EEG signals \cite{rosso2001wavelet}. It was given as
\begin{equation}
    \text{WE} =-\sum_{n=1}^{L}\rho_n \ln(\rho_n),
\end{equation}
where $\rho_n=E_n/E_{\text{total}}$ denotes the relative wavelet energy at the $n^{\mathrm{th}}$ decomposition level, with  $E_n$ and $E_{\text{total}}$ denote the energy of the corresponding wavelet subband, and total signal energy, respectively, and $L$ denotes the maximum decomposition level considered in the wavelet decomposition.

In total, $195$ features comprising band power, coherence, and wavelet entropy measures were extracted and subsequently used for statistical analysis to identify discriminative characteristics among the rest, Samatha, and Vipassana states.

\vspace{-0.2cm}
\section{Results and Discussion}
\label{sec:4}
\vspace{-0.1cm}

This section presents the statistical results obtained by contrasting the two meditation practices: Samatha and Vipassana. Our aim is to investigate whether EEG signals can be used to objectively characterize
Samatha and Vipassana meditation states. Therefore, it is of crucial importance, foremost, that we evaluate the meditation states against the rest state/condition as a comparative baseline. 


\begin{figure}
    \centering
    \includegraphics[width=0.8\linewidth]{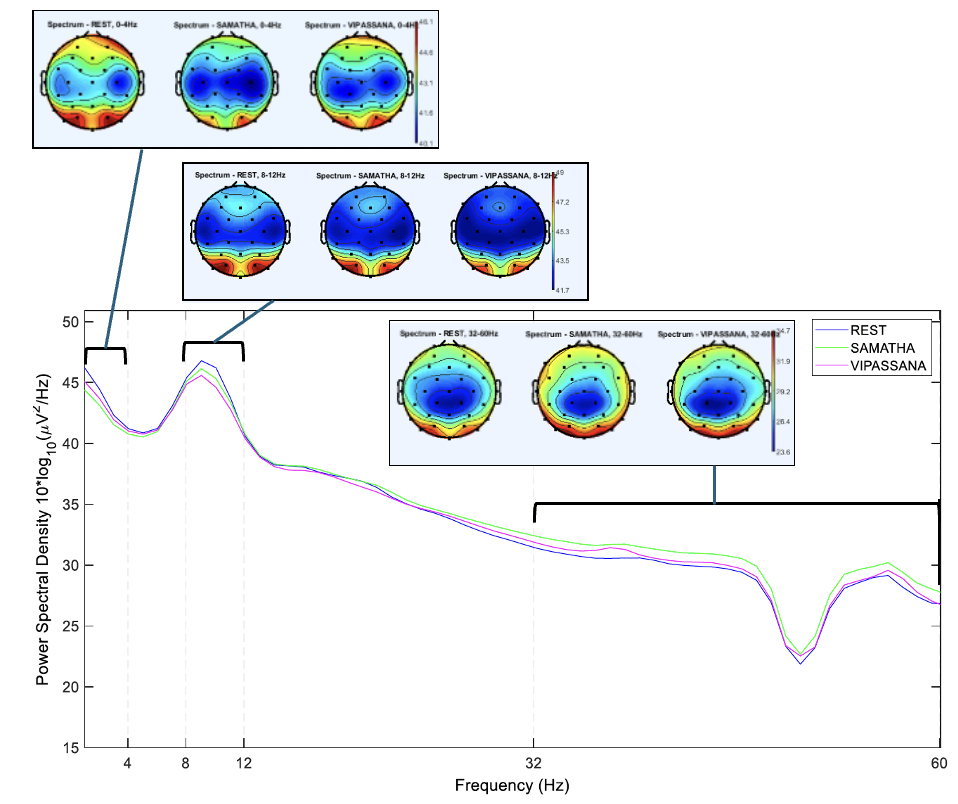}
      \vspace{-0.4cm}
    \caption{The power spectrum average across all electrodes of all the subjects for i) rest, ii) Samatha, and iii) Vipassana meditation states.  Scalp topography of the delta, alpha, and gamma bands across all three states is displayed.}
    \label{fig:realrecsetup}
\end{figure}

\vspace{-0.2cm}
\subsection{{Band Power Analysis}
}\label{sec:4.1}
\vspace{-0.2cm}

Fig.~\ref{fig:realrecsetup} shows the mean amplitude spectral data averaged across all electrodes for rest, Samatha, and Vipassana meditation states with grand average scalp maps for each major EEG bands represented by delta ($0.5 – 4$ Hz), alpha ($8 – 12$ Hz), and gamma ($32 – 60$ Hz). It can be observed that both meditation states are associated with a decrease in overall delta power and alpha power, together with an increase in gamma power, when compared to the pre-meditation resting state. In contrast, no noticeable state-dependent variations were observed in the theta ($4 – 8$ Hz) and beta ($12 – 25$ Hz) frequency bands. These findings suggest that the most prominent spectral changes associated with meditation occur within the delta, alpha, and gamma frequency ranges. Scalp maps for the three states are thus only shown separately for delta, alpha, and gamma frequency bands.

\begin{figure}
    \centering
        \includegraphics[width=0.9\linewidth]{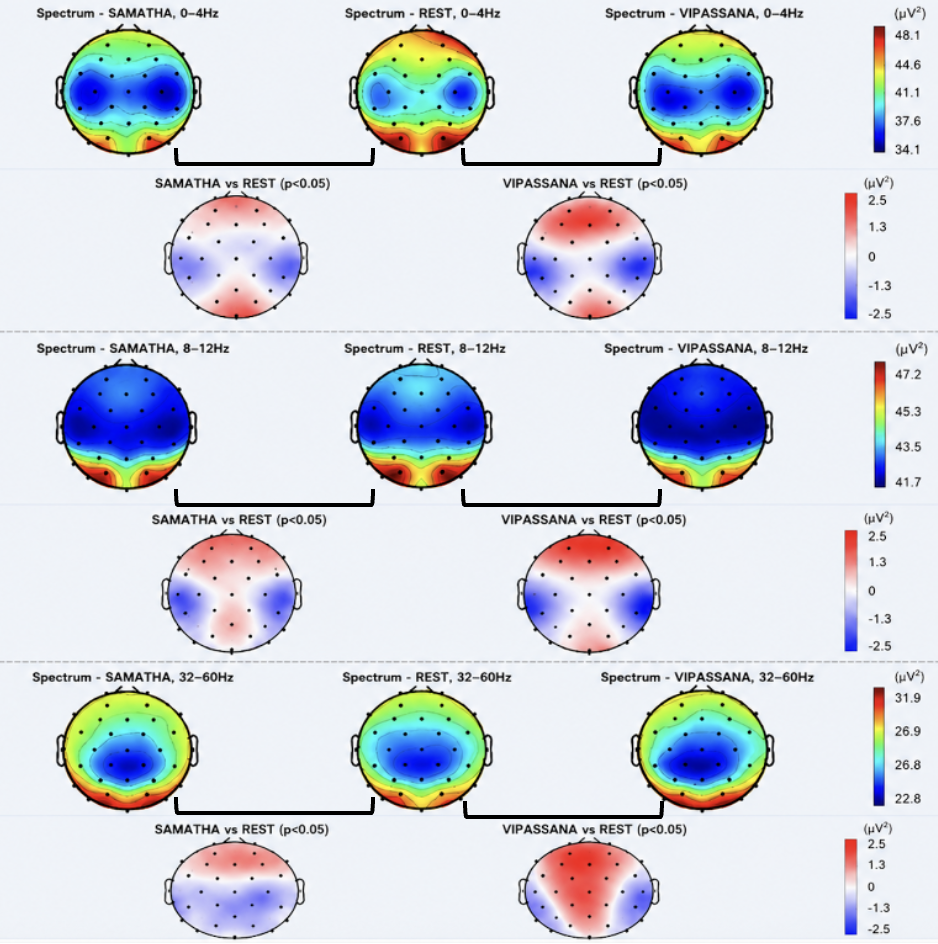}
          \vspace{-0.3cm}
    \caption{The scalp maps for delta ($0.5 – 4$ Hz), alpha ($8 - 12$ Hz), and gamma ($32 – 60$ Hz) bands are displayed with meditation states: $\{$Samatha,Vipassana$\}$, and rest condition in the middle. All three bands showed a statistically significant difference comparing meditation states and rest states, and the scalp map indicating statistical significance is shown below. Statistical significance was determined using a paired T-test with the threshold significance set as $ p< 0.05$.}
    \label{fig:stats}
\end{figure}

Fig.~\ref{fig:stats} presents the spectral power grand average scalp maps for the resting, Samatha, and Vipassana meditation conditions in delta ($0.5 – 4$ Hz), alpha ($8 – 12$ Hz), and gamma ($32 – 60$ Hz) frequency bands, together with the corresponding statistical difference maps (p $< 0.05$) during both Samatha and Vipassana meditation states relative to the resting baseline. 
In the delta and alpha bands, both meditation practices exhibited a similar topographical pattern characterized by significantly increased power over the frontal and anterior scalp regions and decreased power over the central and posterior regions. These findings suggest a common modulation of low-frequency neural activity during meditation compared with the resting condition. In contrast, the gamma band demonstrated a pronounced increase in high-frequency power concentrated over the frontal and anterior-lateral scalp regions for both meditation practices. Also, the Vipassana meditation state exhibited a more extensive and stronger frontal-central gamma power enhancement than the Samatha meditation state, indicating greater engagement of higher-order cognitive processes associated with sustained awareness meditation. Overall, the statistical scalp maps reveal consistent and significant spectral differences in Samatha versus rest and Vipassana versus rest, across all three frequency bands, highlighting distinct neural signatures associated with meditative practice.

\vspace{-0.2cm}
\subsection{{Coherence Analysis}
}\label{sec:4.2}
\vspace{-0.2cm}

Coherence analysis was performed on both inter- and intra-hemispheric electrode pairs across the delta, theta, alpha, beta, and gamma frequency bands. No distinct differences were observed between Samatha and Vipassana meditation states in the alpha, theta, and gamma frequency bands. However, significant differences were identified in the delta band for the inter-hemispheric frontal (F4, \(p=0.048\)) and occipital (PO8, \(p=0.039\)) regions. Also, Vipassana meditation showed higher EEG synchronization than Samatha meditation in the central region (FC5–FC6) for the delta (\(p=0.049\)) and beta (\(p=0.029\)) bands, and in the occipital region (PO7–PO8) for the delta band (\(p=0.044\)).

\begin{figure}[b]
    \centering
        \includegraphics[width=0.8\linewidth]{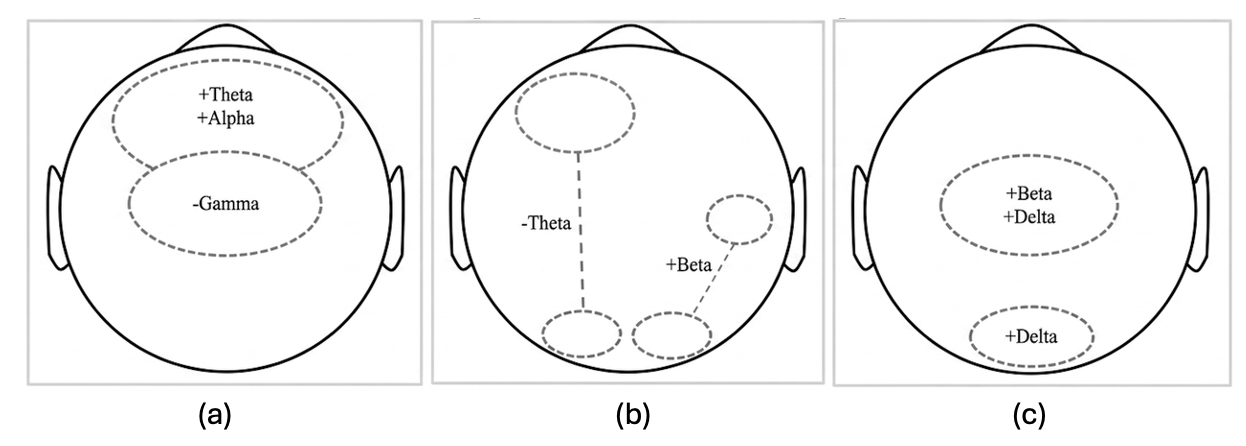}
          \vspace{-0.4cm}
    \caption{Statistically significant coherence features identified between (a) Samatha versus resting states, (b) Vipassana versus resting states, and (c) Samatha versus Vipassana meditation states across the scalp regions. A $`+'$ sign indicates that the corresponding feature is significantly higher in the first meditation state relative to the second, whereas a $`-'$ sign indicates that the feature is significantly lower ($p < 0.05$).}
    \label{fig:coherence}
\end{figure}

When comparing Samatha meditation with the resting state, increased coherence was observed in the frontal region (FP1–FP2) for the alpha (\(p=0.048\)) and theta (\(p=0.036\)) bands, while a significant reduction was found in the central region (CP1–CP2) for the gamma band (\(p=0.027\)). No significant differences were observed in inter-hemispheric coherence between Samatha meditation and the resting state. In contrast, Vipassana meditation exhibited lower theta coherence in the frontal and occipital regions (F3 and PO7, \(p<0.05\)) and higher beta coherence in the temporal and occipital regions (T8 and PO8, \(p<0.05\)) compared with the resting state. The summary of the results obtained from the coherence analysis are depicted in Fig.~\ref{fig:coherence} indicating that coherence changes during meditation are region and frequency-specific, with Vipassana meditation showing stronger synchronization in selected central and occipital regions.

\vspace{-0.2cm}
\subsection{{Entropy Analysis}
}\label{sec:4.3}
\vspace{-0.1cm}

WE was evaluated for the delta, theta, alpha, beta, and gamma frequency bands. Among these bands, the delta band exhibited the largest variation across the resting, Samatha, and Vipassana states, whereas the remaining bands showed only minor differences. Therefore, subsequent WE analysis was performed using the delta frequency band WE.

\begin{figure}[t]
    \centering
        \includegraphics[width=0.7\linewidth]{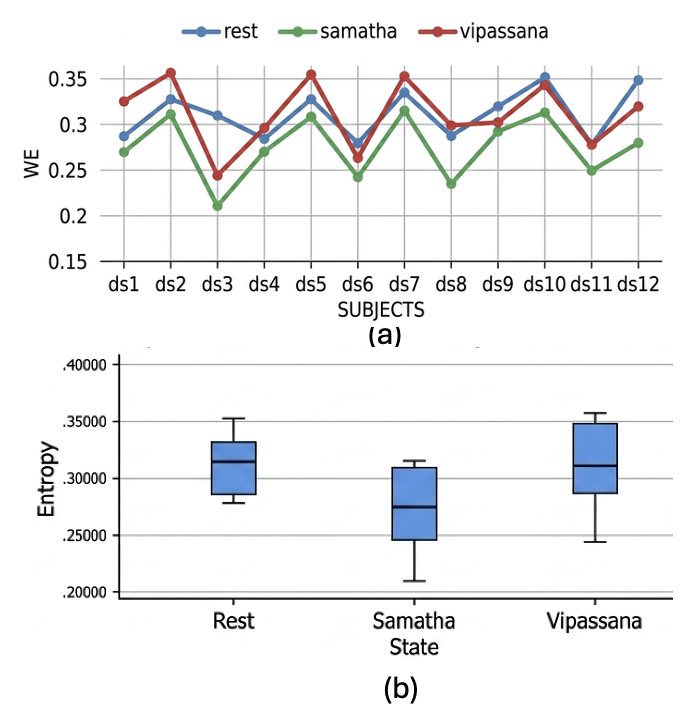}
        \vspace{-0.4cm}
    \caption{Delta sub band WE for the resting, Samatha, and Vipassana meditation states. (a) Average WE across subjects, and (b) corresponding box plot distribution. }
    \label{fig:entropy}
\end{figure}

Fig.~\ref{fig:entropy}(a) shows the average delta band WE for all subjects under the resting, Samatha, and Vipassana conditions across the participants. Compared to the resting state, Samatha meditation consistently resulted in lower WE values across most subjects, indicating a more regular and organized neural activity pattern. In contrast, the WE values observed during Vipassana were generally comparable to those of the resting state, suggesting a similar level of signal complexity.

Fig.~\ref{fig:entropy}(b) depicts the box plot that characterizes the statistical distribution of entropy across the resting, Samatha, and Vipassana states. A comparison of central tendencies revealed that Samatha meditation is associated with a distinct reduction in entropy relative to the resting state. Conversely, the entropy profiles for the resting and Vipassana states demonstrated high similarity, characterized by comparable median values and interquartile ranges. Further analysis indicated statistically significant differences for the Samatha versus rest and Samatha versus Vipassana pairs, however, the difference between the Vipassana and resting states is not statistically significant.

\begin{figure}[t]
    \centering
        \includegraphics[width=0.65\linewidth]{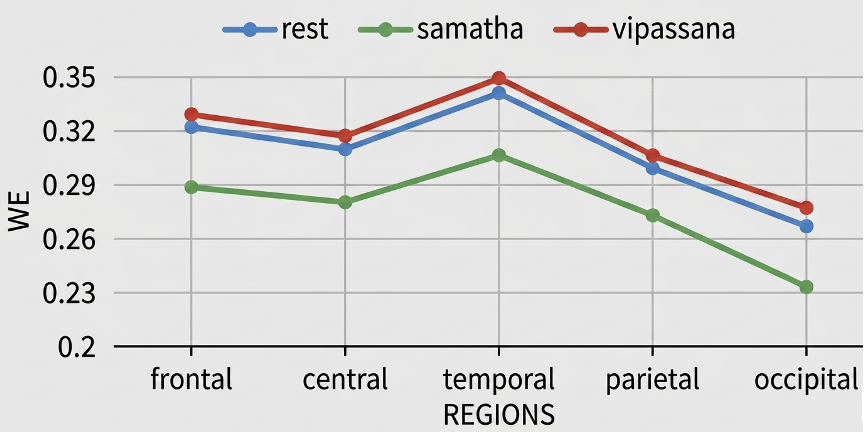}
          \vspace{-0.3cm}
    \caption{Delta frequency band regional WE across scalp regions for the resting, Samatha, and Vipassana meditation states.}
    \label{fig:regionalWE}
\end{figure}

Fig.~\ref{fig:regionalWE} illustrates the regional analysis of WE across the different meditation states. This analysis revealed that Samatha meditation produced lower entropy across all scalp regions, with the largest reduction observed in the occipital region. These findings suggest that Samatha meditation is associated with reduced EEG complexity and increased neural regularity, particularly over occipital cortical areas.

\begin{figure}[t]
    \centering
        \includegraphics[width=0.9\linewidth]{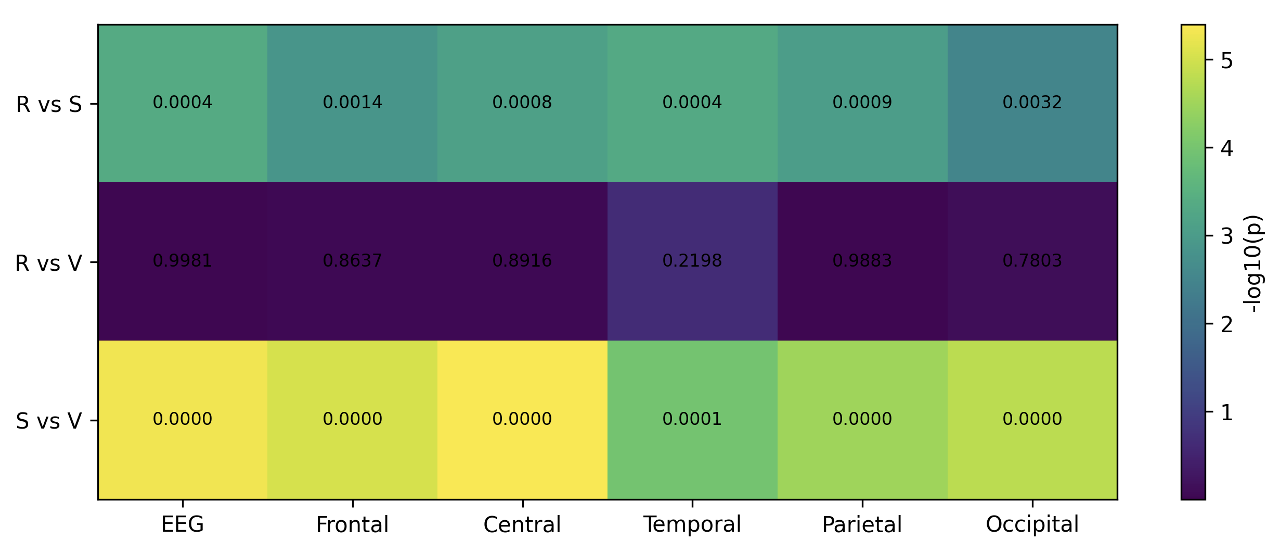}
        \vspace{-0.4cm}
    \caption{Heatmap of p-values obtained from paired t-tests of WE features across EEG regions and overall EEG activity. Here, R, S, and V denote the resting, Samatha, and Vipassana states, respectively. Statistical significance was assessed at $p < 0.05$, with lower p-values indicating greater discrimination between states.}
    \label{fig:colormap}
\end{figure}

Fig.~\ref{fig:colormap} shows a heatmap illustrating the statistical significance of the WE differences among the resting (R), Samatha (S), and Vipassana (V) states across the five scalp regions and the overall EEG activity. Significant differences were observed between R and S in all regions ($p < 0.004$), indicating that wavelet entropy can effectively distinguish the resting state from Samatha meditation. The strongest separation was observed in the temporal region ($p = 4.46 \times 10^{-4}$). Similarly, highly significant differences were found between S and V across all regions ($p < 0.001$), with the central region exhibiting the greatest discrimination ($p = 4 \times 10^{-6}$). In contrast, no significant differences were observed between R and V ($p > 0.05$) for any region or the overall EEG activity. These results suggest that WE effectively differentiates Samatha meditation from both resting and Vipassana states, whereas the resting and Vipassana conditions exhibit similar entropy characteristics.



\vspace{-0.2cm}
\subsection{{Comparison of Vipassana versus Samatha 
}
}\label{sec:4.4}


\begin{figure}[ht]
    \centering
        \includegraphics[width=0.8\linewidth]{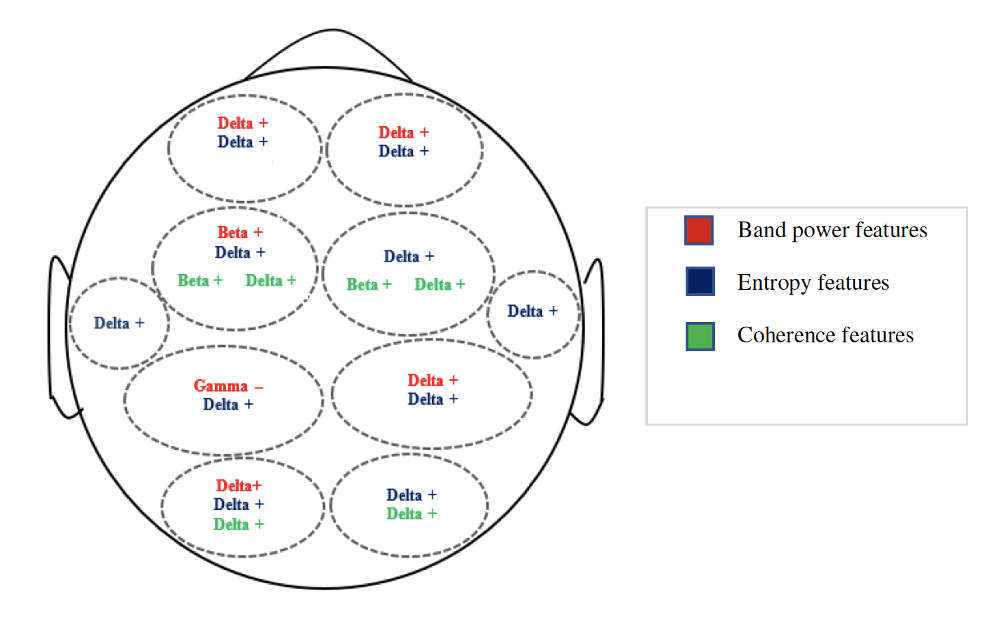}  
          \vspace{-0.6cm}
    \caption{Significant band power, wavelet entropy, and coherence features identified between the Vipassana and Samatha meditation states. A $`+'$ sign indicates a significantly higher feature value in Vipassana relative to Samatha, whereas a $`-'$ sign indicates a significantly lower feature value ($p < 0.05$). }
    \label{fig:svsv}
\end{figure}

In Fig.~\ref{fig:svsv}, we summarized the statistically significant EEG features that differentiate Vipassana and Samatha meditation states with band power, entropy, and coherence, while $`+'$ and $`-'$ symbols indicate whether the corresponding feature is significantly increased or decreased in Vipassana relative to Samatha. It can be seen that the most prominent differences between the two meditation states occur in the delta band. Significant increases in delta band power, entropy, and coherence are observed across multiple frontal, central, parietal, occipital, and temporal regions, indicating that low frequency neural activity and connectivity are more pronounced during Vipassana meditation. In contrast, localized reductions in delta band entropy are observed in the temporal region, suggesting regional differences in signal complexity between the two meditation practices.

Moreover, significant beta frequency band power and coherence increases are observed in the frontal and central regions, indicating enhanced synchronization and cortical engagement during Vipassana meditation. A significant gamma frequency band power reduction is evident in the left temporal region, suggesting lower high frequency activity in Vipassana compared to the Samatha within this localized cortical area.

Overall, Fig.~\ref{fig:svsv} demonstrated that delta band features consistently emerge across all three analysis domains: band power, entropy, and coherence, across several scalp regions. 
These findings are consistent with the underlying nature of the two meditation practices, where Samatha emphasizes calmness, concentration, and mental stability, whereas Vipassana involves active observation and analysis of ongoing mental processes \cite{cahn2010occipital}. Consequently, Vipassana exhibits relatively higher neural complexity and connectivity, while Samatha reflects a more relaxed and organized cognitive state.


A few comments on the implementation procedures of the study:
\begin{itemize}
    \item{Age variability: although age may influence general EEG characteristics, 
    the attainment of advanced meditation states
    is primarily related to meditation proficiency rather than chronological age. Therefore, age variability is not expected to directly affect the investigated meditation-related EEG patterns.}
    
    \item{Meditation order effect: the meditation order was not randomized, as Samatha was performed before Vipassana following traditional Buddhist practice. A potential order effect cannot be excluded and will be investigated in future studies using randomized protocols.}
\end{itemize}

\vspace{-0.2cm}
\section{Conclusions}
\label{sec:5}
\vspace{-0.1cm}

This study aims at investigating whether EEG signals can be used to characterize Samatha and Vipassana meditation practices based on EEG recordings acquired during meditation sessions. The work focused primarily on the delta, alpha, and gamma frequency bands using band power, coherence, and wavelet entropy analysis. The preliminary results revealed several significant features and corresponding scalp regions that distinguish meditation states from the pre-meditation resting condition, as well as differentiate between Samatha and Vipassana meditation practices. Overall, the findings indicate that both forms of meditation are associated with measurable changes in brain dynamics, reflected through alterations in spectral power, signal complexity, and functional connectivity. These observations provide further insight into the neural processes underlying meditation and suggest that EEG-based measures may serve as useful tools for objectively assessing and distinguishing different meditation states.

\vspace{-0.2cm}
\section*{Acknowledgment}
The authors would like to thank Dr. Kamal Gunarathna, Consultant Clinical Neurophysiologist at the National Hospital of Sri Lanka, and Dr. Anjula De Silva for their valuable support and guidance throughout this work. 











\printbibliography

\end{document}